\documentclass[journal = jacsat, manuscript = communication, layout = twocolumn]{achemso}

\usepackage[version=3]{mhchem}
\usepackage{xcolor}
\usepackage{amsmath}
\usepackage{amsfonts}

\author{Linda A. Zotti}
\email{linda.zotti@uam.es}
\affiliation[Universidad Aut\'onoma de Madrid]{Departamento de F\'{\i}sica de la 
Materia Condensada, Universidad Aut\'onoma de Madrid, 28049 Madrid, Spain}

\author{Edmund Leary}
\affiliation[Universidad Aut\'onoma de Madrid]{Departamento de F\'{\i}sica de 
la Materia Condensada, Universidad Aut\'onoma de Madrid, 28049 Madrid, Spain}
\alsoaffiliation{Inst Madrileno Estudios Avanzados Nanociencia IMDEA, E-28049 Madrid, Spain}

\author{Maria Soriano}
\affiliation[Universidad Aut\'onoma de Madrid]{Departamento de F\'{\i}sica de la 
Materia Condensada, Universidad Aut\'onoma de Madrid, 28049 Madrid, Spain}
\alsoaffiliation[Universidad de Alicante]{Departamento de F\'{\i}sica Aplicada, Universidad de Alicante, 03690 Alicante, Spain}

\author{Juan Carlos Cuevas}
\affiliation[Universidad Aut\'onoma de Madrid]{Departamento de F\'{\i}sica Te\'orica 
de la Materia Condensada, Universidad Aut\'onoma de Madrid, E-28049 Madrid, Spain}

\author{Juan Jose Palacios}
\affiliation[Universidad Aut\'onoma de Madrid]{Departamento de F\'{\i}sica de la 
Materia Condensada, Universidad Aut\'onoma de Madrid, 28049 Madrid, Spain}

\title{A Molecular Platinum Cluster Junction: A Single-Molecule Switch}

\begin{document}

\begin{abstract}
 We present a theoretical study of the electronic transport through
 single-molecule junctions incorporating a Pt$_{6}$ metal cluster bound within 
 an organic framework. We show that the insertion of this molecule between
 a pair of electrodes leads to a fully atomically-engineered nano-metallic
 device with high conductance at the Fermi level and two sequential
 high on/off switching states. The origin of this property can be traced back 
 to the existence of a HOMO which consists of two degenerate and asymmetric 
 orbitals,  lying close in energy to the Fermi level of the metallic leads.
 Their degeneracy is broken when the molecule is contacted to the leads, giving
 rise to two resonances which become pinned close to the Fermi level and display 
 destructive interference.
\end{abstract}

Molecular metal clusters have recently started  being explored in breakjunction 
type single-molecule-conductance experiments \cite{vanderZant2006,Leary,Venkataraman},whilst previous studies have also investigated clusters in the STM-STS configuration 
which mantains a tunnel gap between the tip and the molecule. \cite{Soldatov,Gubin}
Clusters have fewer metal atoms than metal nanoparticles but a more precisely 
defined composition and structure and as such they are better suited for 
comparative studies between theory and experiments. Bare metal clusters have 
been successfully deposited and studied on various surfaces.\cite{Yasumatsu,Isomura} 
Such clusters however require very clean conditions for study due to their high 
reactivity. In order to overcome this fact, and study molecules under the
normal ambient conditions of the breakjunction experiment,
 the metal atoms must be encapsulated
in a ligand sphere to avoid unwanted chemical reactions with the atmosphere. 
Ligand stabilized metal clusters (in particular those incorporated in an organic 
framework) have therefore  been suggested as components of data storage devices 
where they would act as nano-capacitors due to their redox properties.\cite{Femoni2006}
Transport measurements on individual molecular metal clusters are, however, still 
rare. Metal atoms have been successfully incorporated into organic frameworks
as metal complexes for single-molecule experiments, as well as
chains of atoms.  \cite{Park2002,Ruben2008,Mayor2002,Georgiev} Regarding
metal clusters,  the Mn$_{12}$
structure has been investigated as the functional core in single molecule magnets
\cite{vanderZant2006} and theoretical studies about its transport properties
have been carried out.\cite{Renani2011,Ferrer2009} However, despite the vast 
range of molecular clusters known, only a few have been analyzed in 
this context. This is  especially true regarding the theory of the electronic 
transport.\cite{Gerasimov2010,Gerasimov2011}

In ref. \cite{Leary}, single-molecule experiments on \\
\ce{[Pt_{6}(\mu-P^{t}Bu_{2})_{4}(CO)_{4}(S(CH_{2})_{4}SH)_{2})3]}
(from now on (SC$_{4}$S)$_{2}$Pt$_{6}$) were carried out using the STM based 
I(s) technique \cite{Haiss,Nichols2010}. The  structure of this Pt cluster 
makes it ideal for these kind of measurements as the ligands which act as 
binding groups (the alkanethiol chains) are oriented trans to each other,
 producing a linear wire. 
Without these chains, it would be difficult to determine a
 precise point of attachment between the molecule and the electrodes.
 The presence of these groups not only provides robust thiol anchors,
 but also increases the total length of the molecule in one defined
 direction, allowing measured break-off distances to assess accurately
 the orientation of the molecule in the junction
(the need of incorporating the metal cluster in a wire has 
also been suggested in a theoretical study of the electronic transport
through a Mn$_{12}$ cluster \cite{Ferrer2009}).  
In the experimental study, it was suggested that the 
presence of the platinum cluster increases the conductance compared to an 
alkane chain of equivalent length due to there being Pt states in close 
proximity to the Fermi level. These were claimed to create an indentation 
in the potential barrier, giving a molecular analog of an inorganic 
double tunneling barrier, with the Pt$_{6}$ unit acting as a well and the alkane 
chains acting as barriers so to decouple the central unit from the leads.
This was, however, a speculative argument which was never fully proven, and neither 
was the nature of these states identified. Finding out whether this assertion 
could be true is important because the molecular double tunneling barrier 
configuration could be the basis for creating molecular switches \cite{Leary2007}, 
analogous of conventional electronic components. This is ultimately one of 
the main aims of molecular electronics \cite{Cuevas2010}. In this study 
presented here, we carried out first principle calculations to find out if 
we could corroborate the assumptions made in the experimental work. We will show 
that the HOMO of (SC$_{4}$S)$_{2}$Pt$_{6}$ consists of two degenerate levels 
which  get pinned to the Fermi level when the molecule is placed between 
two gold electrodes. However, these two levels do not originate simply from
the Pt$_{6}$ unit, but contain states localized on the ligands also. Moreover, 
interesting interference features appear in the transmission curve at the 
Fermi level. 

\begin{figure}[t]
\begin{center}
\includegraphics[width=8.0cm]{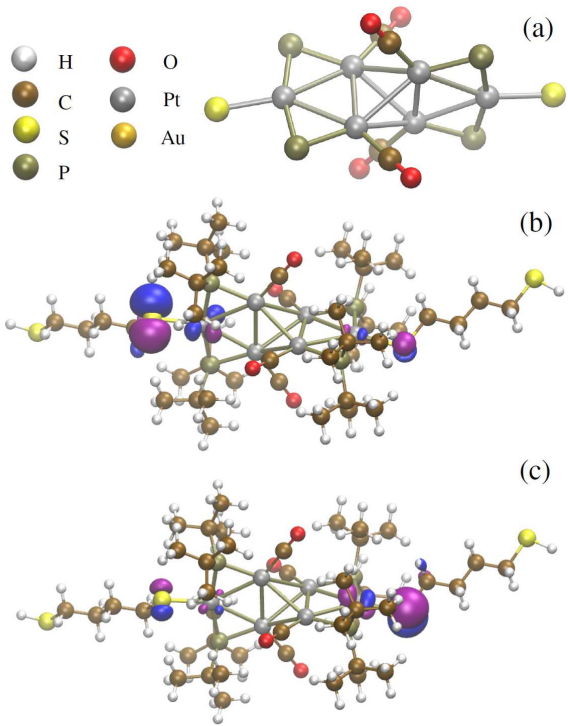}
\caption{ Central moiety of (SC$_{4}$S)$_{2}$Pt$_{6}$ (a) and
optimized geometry of the whole molecule with the two degenerate
HOMO orbitals (b-c). 
} \label{fig:orbitals}
\end{center}
\end{figure}

We optimized all geometries with Turbomole 6.1 \cite{Turbomole}, using the 
BP86 functional \cite{BP86} and the def-SVP basis set \cite{Schafer1992}. 
We built the molecular junctions by placing the molecules (previously relaxed 
in the gas phase) between two 20 gold atom clusters and then relaxed the 
molecule and the 4(3) innermost gold atoms on each side in a top (hollow) 
geometry. The subsequent transport calculations were carried out with ANT
\cite{alacant}, which is built as an interface to Gaussian \cite{Gaussian}. 
This code computes the electronic transmission using  nonequilibrium Green's function
 techniques in the spirit of the Landauer formalism, employing parametrized
 tight-binding Bethe lattices in the electrode description 
\cite{Jacob2011}. For these calculations, we used the PBE  functional \cite{PBE},
a lanl2DZ basis set for Pt atoms and the four innermost Au atoms on each side 
\cite{lan2dz}, while we used a CRENBS basis set \cite{crenbs} for all the 
other Au atoms. A 6-31++G basis set was chosen for C, S, P, O, and sto-3g for H. 

We first studied the molecule in the gas phase. The optimized geometry is
shown in \ref{fig:orbitals}. It consists of a Pt$_{6}$ cluster
sandwiched between two butanedithiol (four carbon atoms) chains.
 The central cluster consists in turn
of two orthogonal Pt$_{3}$ triangles, confirmed by crystallographic structural
 determination  on the carbonyl substituted compound \cite{deBiani}.
Each Pt$_{3}$ triangle contains two bridging phosphine groups and the 
remaining sides are joined together to form  a Pt tetrahedral cored.
The tetrahedral core also contains four CO ligands.
The molecule is slightly bent, with the angle between the alkyl chains
less than 180$^{\rm o}$. In the second and third panel of \ref{fig:orbitals}
we show the HOMO. It consists of two doubly occupied levels which are
 degenerate and the corresponding orbitals of which
are related to each other by a 4-fold improper rotation.
 It is mainly localized on the  two innermost S atoms and on the two
neighboring Pt atoms.
Notice that these frontier orbitals are different from the HOMO shown
in ref \cite{deBiani} due to the presence of the butanedithiol ligands.
We also studied a C$_{18}$H$_{36}$ alkyl chain terminated with S atoms
 (from now on C$_{18}$),
 as it was claimed \cite{Leary} to have the same length as
 (SC$_{4}$S)$_{2}$Pt$_{6}$ but 
a much lower conductance. Such a comparison would help us decide whether
the presence of the Pt cluster provides a significant change in the
 conductance with respect to the alkyl chain alone.
 We first checked that the length of this chain provides a reasonable
 comparison to the cluster molecule. The length of C$_{18}$
 is 24.9 \AA{}, which is halfway between the through space S to S distance
 (23.9 \AA{}) and the length measured by adding the 3 components
 (2 chains + Pt core, 25.8 \AA{} ) in (SC$_{4}$S)$_{2}$Pt$_{6}$ . 
In C$_{18}$, the HOMO  is localized on the S atoms, 
while the HOMO-1 is delocalized throughout the whole chain \cite{Cuevas2010} .

\begin{figure}[t]
\begin{center}
\includegraphics[width=7.0cm]{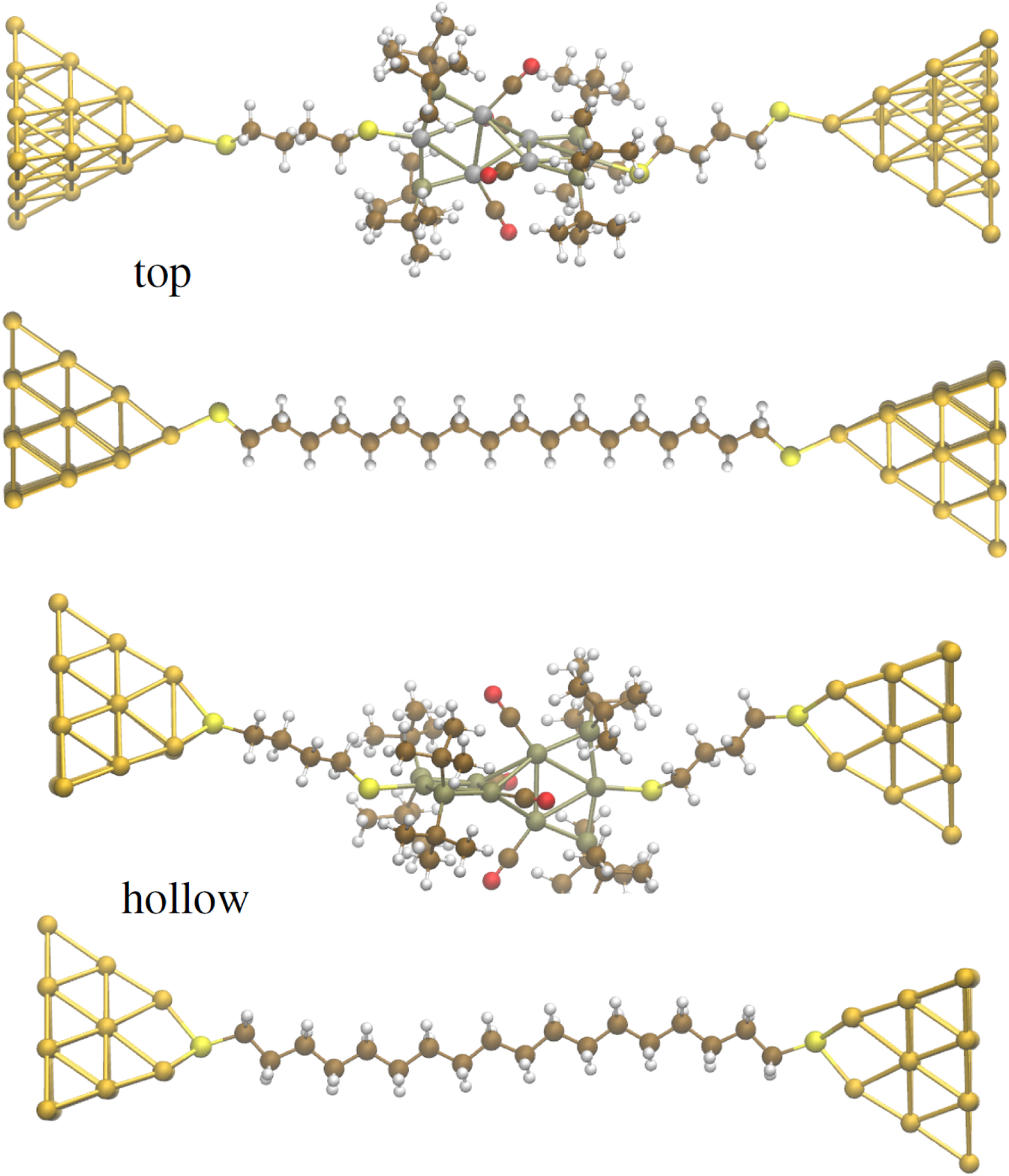}
\caption{(Color online) Optimized geometries for (SC$_{4}$S)$_{2}$Pt$_{6}$
 and C$_{18}$ embedded
between Au clusters.
} \label{fig:junctions}
\end{center}
\end{figure}

\begin{figure}[t]
\begin{center}
\includegraphics[width=8.0cm]{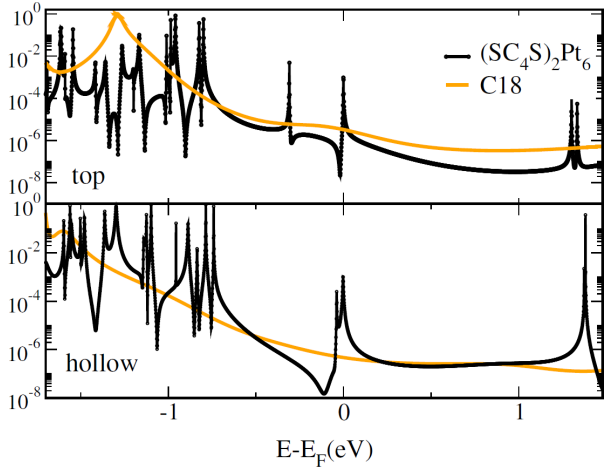}
\caption{Transmission as a function of energy for all studied molecular junctions.
} \label{fig:transmiss}
\end{center}
\end{figure}

After analyzing the molecules in the gas phase, we proceeded to study
junctions containing (SC$_{4}$S)$_{2}$Pt$_{6}$ and C$_{18}$
 bound to the gold
clusters in a top and a hollow geometry (\ref{fig:junctions}).
 \ref{fig:transmiss} shows the transmission curves for all four cases.
In the case of C$_{18}$, the energy alignment seems to be considerably
affected by the binding geometry. In the hollow geometry, the HOMO-1
(which is delocalized throughout the whole chain and gives the main 
contribution  to the transmission \cite{Cuevas2010})
 is shifted down in energy compared to the top position
 (as  has also been observed 
 for other thiolated molecules \cite{Zotti2010}), yielding a difference
in the conductance of one order of magnitude between the two geometries.
Notice also the presence of the HOMO (localized on the S atoms) very close
to the Fermi level, clearly visible as a bump in the transmission curve
for the top geometry.
In the case of (SC$_{4}$S)$_{2}$Pt$_{6}$,
the alignment of the HOMO states does not show a particular dependence
on the binding geometry as they are pinned at the Fermi
level in both cases, but rather their splitting seems to be affected.
The degeneracy of the HOMO states is broken in
the junction as the symmetry has changed upon geometry relaxation.
We stress that
the two peaks close to the Fermi level do not arise simply from the Pt$_{6}$ unit
as suggested in ref. \cite{Leary}, but rather from orbitals which contain both
S and Pt contribution,  which however do not spread over all Pt atoms,
as discussed above. 

Fermi level pinning has been widely
studied  but its nature is still under discussion
(see ref \cite{Bokdam2011} and references therein).
It is also well known that molecular HOMO-LUMO gaps are usually underestimated
 in DFT and that this affects the conductance values in DFT-based transmission
 calculations.
In order to gain insight into the level alignment in our calculations
and to understand whether it is reliable, we evaluated
the ionization potential (IP) and electron affinity (EA) 
in the gas phase for the two molecules, as $E(N)-E(N-1)$ and $E(N+1)-E(N)$,
respectively, where $N$ is the total electron number.
These quantities (based on total-energy differences) are expected to be more
reliable than the HOMO and LUMO
assigned by the Kohn-Sham states (at -4.27  and -3.07 eV for
 (SC$_{4}$S)$_{2}$Pt$_{6}$
and at -5.3 and -0.08 eV for C$_{18}$, respectively).
The calculated IP and EA in the gas phase were found
 to be at -5.66 and -1.70 eV  for (SC$_{4}$S)$_{2}$Pt$_{6}$, and -7.4 and 0.45
 eV for C$_{18}$, respectively.
Molecular HOMO-LUMO gaps are expected to narrow when molecules approach
 metal electrodes due to the screening effect \cite{Thygesen2009}. 
In the case of (SC$_{4}$S)$_{2}$Pt$_{6}$, since the IP is very close to
 the Au Fermi level (which is at -5.0 eV),
 it is plausible that contacting the molecule to the electrode raises the IP
until such a point that it becomes pinned to the Fermi level. This is
accompanied by a charge transfer from the molecule onto the metal,
 confirmed by 1.47 and 4.3 e positive charge found on the
 molecule in the top   and  hollow position, respectively. Notice
that  the transferred charge originates mostly from the S atoms
directly connected to the leads.
Based on the proximity of the IP to the gold Fermi level, we believe
 that the energy alignment of the HOMO states appearing in our
 transmission curves for (SC$_{4}$S)$_{2}$Pt$_{6}$
is robust and  unaffected by common DFT failures, providing a realistic
picture of the experimental scenario. On the other hand,
 in the case of C$_{18}$, the IP calculated in the gas phase is at
a much lower energy than  the peak corresponding to the HOMO
 (localized on the S atoms) in the transmission curve for the
 Au-C$_{18}$-Au junction, especially for the top binding geometry.
This suggests that it should align at lower energy; if so,
 the conductance values of C$_{18}$ would be significantly lower
 than those which we computed. 

\begin{figure}[t]
\begin{center}
\includegraphics[width=8.0cm]{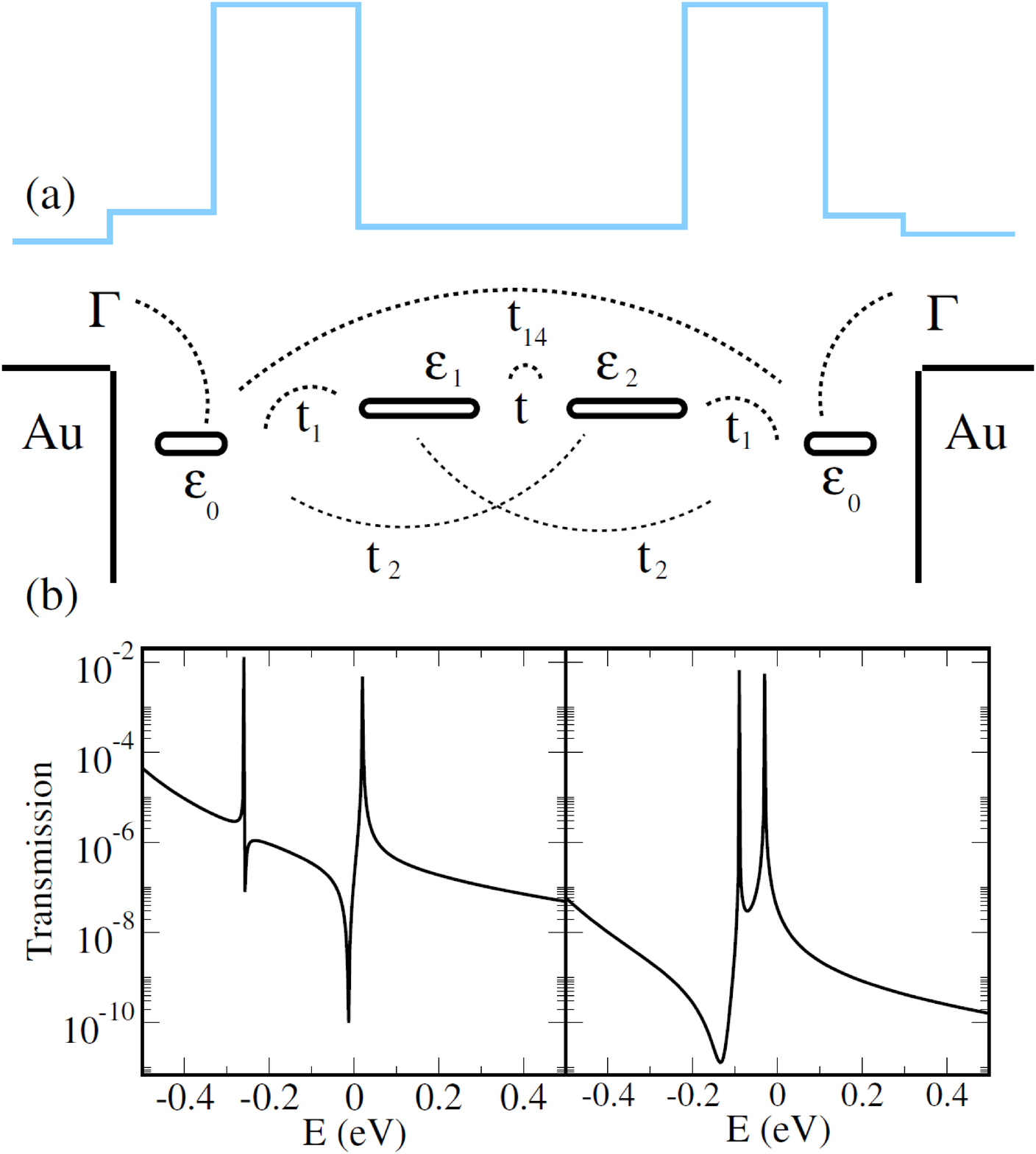}
\caption{(a) Schematic representation (black) of the toy model describing the
interference effects close to the Fermi energy and of the double
tunnel barrier (blue). (b) The corresponding 
transmission as a function of energy for the case of the top position (left
panel) with the following parameter values: $\varGamma=0.05$ eV, 
$\varepsilon_0=-0.7$ eV, $\varepsilon_1 = \varepsilon_2= -0.12$ eV, 
$t_1=0.009$ eV, $t_2=0.005$ eV, $t=0.14$ eV, $t_{14}=0.003$ eV, and for the
hollow geometry (right panel) with the parameter values: $\varGamma=0.05$ eV,
$\varepsilon_0=-0.7$ eV, $\varepsilon_1=\varepsilon_2= -0.06$ eV, $t_1=0.005$
eV, $t_2=0.002$ eV, $t=0.03$ eV, $t_{14}=0.000145$ eV.}
\label{fig:draw-xfig}
\end{center}
\end{figure}

A further significant result is the presence of interference resonances
 in the transmission
 curves, which are clearly visible in \ref{fig:transmiss}.
The interest in interference effects in molecular junctions is
 steadily growing
 \cite{Bergfield2011,Markussen2010,Cuniberti2012,Lambert2012}
 and this stems from the fact that they
modify the thermoelectric properties of the junctions
 \cite{Cuevas2010} and that the ensuing dips in the transmission
 curves could give  rise to large on/off ratios
in future molecule-based electronic devices \cite{Cuniberti2012}.
Interference effects have recently been detected experimentally
in molecular junctions \cite{vanderMolen2012}.
In our case, they originate from the existence of two possible pathways
 (the two HOMO states).
 For the top position, we find  a similar feature to that produced
 in ref. \cite{Satanin2006}
 Interferences yielding peaks at the Fermi level have already been
predicted \cite{Finch2009, Wei2012}, however their occurrence from
 two degenerate levels pinned at the Fermi level, to the best
of our knowledge, is rare. 
In order to shed light on the origin of these interference features,
we built a simple model as depicted in \ref{fig:draw-xfig}, where
we consider two levels both connected to the leads.
They are related to the two HOMO levels of (SC$_{4}$S)$_{2}$Pt$_{6}$,
which are degenerate and orthogonal in the gas phase.  However,
when the molecule is placed between two electrodes, the symmetry
and consequently the degeneracy is broken due to the geometrical
readjustment of the molecule. Thus, in our model, we consider two
states which are no longer eigenstates of the system and are coupled
by a non-zero hopping matrix element $t$.

In this model, the Hamiltonian of the central (molecular) part takes the 
following form:
\[H_C= \left(\begin{array}{cccc}
\varepsilon_0 & t_2 & t_1 & t_{14} \\
t_2 & \varepsilon_2 & t & t_{1}\\
t_{1} & t & \varepsilon_1 & t_{2} \\
t_{14} & t_{1} & t_{2} & \varepsilon_0 \end{array}\right)\]
where $\varepsilon_1$  and $\varepsilon_2$ are the energies of two degenerate
molecular levels, coupled to each other via $t$. $\varepsilon_0$ is the energy 
of the interface levels, coupled
to $\varepsilon_1$  and $\varepsilon_2$  through $t_1$ and $t_2$.
Notice that $t_1$ must be bigger than $t_2$, in order to account for
the asymmetry of each orbital.
The transmission is given by
\begin{equation}
 T(E) = 4 \varGamma_L \varGamma_R |G_{14}|^2
\end{equation}
where
\begin{equation}
 G=[E-H_C-\varSigma]^{-1}
\end{equation}
and $\varSigma=\varSigma_L+\varSigma_R$, being $\varSigma_L$
and $\varSigma_R$ the self energies for left and right lead, respectively.
These two matrices have all terms equal to 0 but $\varSigma_{L11}=i\varGamma_L$
and $\varSigma_{R44} = i \varGamma_R$, being $\varGamma_L$ and $\varGamma_R$
the coupling to left and right lead, respectively. In our case, $\varGamma_L =
\varGamma_R =\varGamma$ as we are considering symmetric junctions.

As an example, in the inset of \ref{fig:draw-xfig} we show
 the transmission curve given by this model for a chosen set of
 parameters (listed in the caption) which can reproduce the shape
of the DFT-calculated transmission curve for both hollow and top position.
This proves that the nature of the  features appearing
in the transmission curves at the Fermi level is in the interference
between the two levels of the central moiety  and two levels localized
 at the interface between this central part and the leads.
In the case of asymmetric junctions, such as top-hollow, the transmission
curve is expected to resemble that corresponding to the top-top case
(see the Supporting Information).

 As mentioned above, the presence of a resonance immediately
 followed by an antiresonance increases the on/off ratio
 in gating experiments. One can simply view this by imagining
 that the Fermi level (located at 0 eV) is shifted to more negative values along
 the energy axis: by doing this, in the top geometry, for instance,
  the Fermi level will cross a first sharp peak (\emph{on})
 and, after a very small shift, a sharp dip (\emph{off}).
  This sequence is then
 repeated by shifting the Fermi level further so as to cross
 the second resonance-antiresonance couple. Indeed, we propose
that the low bias conductance of this system should be measured
 experimentally as a function of the gate voltage,
 in order to test the breaking of the degeneracy.

Finally, we compare our results with the experiments.
In ref. \cite{Leary}, a conductance of $3 \times 10^{-5} G_{0}$ was 
measured for (SC$_{4}$S)$_{2}$Pt$_{6}$, while the conductance
of C$_{18}$ was extrapolated from the experimental attenuation (beta) 
value measured for a series of shorter compounds as $6 \times 10^{-10} G_{0}$.
It was argued that the alkyl chains
 in (SC$_{4}$S)$_{2}$Pt$_{6}$ simply act  as
 spacers as the frontier orbitals lie far from the Fermi level,
 while the frontier
orbitals of the  central unit create a barrier indentation and,
 consequently, raise the conductance by approximately  6 orders 
of magnitude compared to C$_{18}$.
In our results, it is true that precisely at the very Fermi level
 the conductance of (SC$_{4}$S)$_{2}$Pt$_{6}$  is definitely higher
 than C$_{18}$; however, at other energies
immediately below or above it, the conductance is comparable or even lower,
especially as a result of the destructive interference. 
This could potentially explain some non-linearities observed in some of the I-V
curves recorded experimentally (see the Supporting Information of
ref \cite{Leary}). 
It is also worth adding that, due to the uncertainty about
the precise length of the Pt molecule, exactly comparing the
conductances of (SC$_{4}$S)$_{2}$Pt$_{6}$ and C$_{18}$ is somehow
ambiguous, since, in our equilibrium geometries, C$_{18}$ is straight
(with only slight defects) whereas the cluster molecule is  bent, 
with the Pt$_{6}$ cluster slightly out of the Au-Au axis. 
 In the experiments though, the Pt molecule is probably straighten,
due to the pulling stress applied. 
Regardless these uncertainties, the overall picture supports
 the naive idea based on the experimental finding, confirming
the presence of resonances at the Fermi level arising from the presence
of the central additional moiety in the molecular cluster. 

In conclusion, we have theoretically studied single molecule junctions
 incorporating a Pt$_{6}$ cluster, showing that it acts in an analogous
 fashion to a double tunneling barrier due to two (originally degenerate)
 states which align at the Fermi level. These states do not, however, stem
 from the Pt unit alone, but specifically from two apical Pt atoms and
 their neighboring S atoms. This gives rise to quantum interference effects
 due to multiple electronic pathways through the molecule.
Due to the pinning, these effects should be easily detectable experimentally.
 Despite the seemingly  apparent complexity of the structure,
 all the physical properties arise from these four atoms alone, highlighting
 the delicate relationship between molecular structure and electrical properties.
 We propose that further chemical synthesis combined
 with theoretical calculations should be performed in order to explore more
 combinations of metal clusters with organic moieties. This could provide
 an alternate design strategy for
 molecular devices other than through purely organic molecules due to the
 wealth of different structures possible via inorganic chemistry.
Recent progress in implementing a gate in single-molecule based devices
 \cite{Park2000, Zhang2008,Leary2008,vanderZant2011} allows us to predict
 for the studied molecule,  with its controlled binding to the electrodes
 and with its two levels aligned at the Fermi level,  that it could be used
 in a three-terminal device in which only a small gate voltage would be
 enough to pass through two high/low sequential switching states.

\begin{acknowledgement}
This research was supported by the Comunidad de Madrid through the project
NANOBIOMAGNET S2009/MAT1726, by the Generalitat Valenciana through the
project PROMETEO2012/011 and by the Spanish MICINN under the Grant Nos.
 FIS2010-21883 and CONSOLIDER CSD2007-0010. 
EL was funded by the EU through the ELFOS Network (FP7-ICT2009-6).
We thank the CCC of Universidad Aut\'onoma de Madrid for computational
resources.
\end{acknowledgement}

\bibliography{Literature_Zotti_150113}

\end{document}